\documentclass[sigconf,screen]{acmart}
\usepackage{graphicx} % Required for inserting images
\usepackage{xspace}
\usepackage{url}
\usepackage{array}
\usepackage{makecell}
\usepackage{tabularx}
\usepackage{multirow}
\usepackage{fontawesome5}
\newcommand{\tool}{\textsc{\textbf{SmellCC}}\xspace}

\AtBeginDocument{%
  }

\setcopyright{cc}
\setcctype{by}
\acmDOI{10.1145/3832783.3834611}
\acmYear{2026}
\copyrightyear{2026}
\acmISBN{979-8-4007-2882-2/2026/10}
\acmConference[ASE '26]{Proceedings of the 41st IEEE/ACM International Conference on Automated Software Engineering}{October 12--16, 2026}{Munich, Germany}
\acmBooktitle{Proceedings of the 41st IEEE/ACM International Conference on Automated Software Engineering (ASE '26), October 12--16, 2026, Munich, Germany}
\acmSubmissionID{ase26tool-p47-p}
\received{2026-05-11}
\received[accepted]{2026-06-19}

\begin{document}

\title{\textsc{\textbf{SmellCC}}: A Tool for Automated Code Smells Remediation}

\author{Xiaoting Zhang}
\affiliation{%
  \institution{Zhejiang University, \\ Hangzhou High-Tech Zone (Binjiang) Institute of Blockchain and Data Security}
  \city{HangZhou}
  \country{China}
}
\email{xiaotingzhang@zju.edu.cn}

\author{Yujie Zhang}
\affiliation{%
  \institution{Zhejiang University, \\ Hangzhou High-Tech Zone (Binjiang) Institute of Blockchain and Data Security}
  \city{HangZhou}
  \country{China}
}
\email{zhang.yujie@zju.edu.cn}

\author{Zhipeng Gao}
\authornote{Corresponding author.}
\affiliation{%
  \institution{Zhejiang University, \\ Hangzhou High-Tech Zone (Binjiang) Institute of Blockchain and Data Security}
  \city{Hangzhou}
  \country{China}
}
\email{zhipeng.gao@zju.edu.cn}

\author{Xing Hu}
\affiliation{%
  \institution{Zhejiang University, \\ Hangzhou High-Tech Zone (Binjiang) Institute of Blockchain and Data Security}
  \city{HangZhou}
  \country{China}
}
\email{xinghu@zju.edu.cn}

\author{Xin Xia}
\affiliation{%
  \institution{Zhejiang University, \\ Hangzhou High-Tech Zone (Binjiang) Institute of Blockchain and Data Security}
  \city{HangZhou}
  \country{China}
}
\email{xin.xia@acm.org}


\begin{abstract}
Code smells significantly threaten software maintainability by accumulating technical debt, yet developers often lack the resources to manually address these flaws under tight release schedules. 
While static analysis tools like SonarQube provide precise detection, they function largely as passive alert systems, leaving the burden of refactoring on developers. 
To bridge this gap, we present a novel cleaning tool, namely \tool (\textbf{\underline{Smell}} \textbf{\underline{C}}ode \textbf{\underline{C}}leaner), a Visual Studio Code extension that augments SonarQube with an LLM-based pipeline to automatically detect and refactor Python code smells. 
By employing Chain-of-Thought (CoT) and few-shot learning, \tool provides in-place, one-click remediation for the top-10 most frequent smells, effectively preventing the accumulation of technical debt during development.
Our quantitative evaluation demonstrates that our \tool is promising in helping developers effectively eliminate code smells (96.8\% cleaning rate) with high accuracy (i.e., 91.3\%), ensuring that the refactored code remains syntactically correct and behavior-preserving, thereby significantly improving long-term software maintainability. 
The full paper underlying this tool has been published in ACM Transactions on Software Engineering and Methodology~\cite{10.1145/3793252}, where \tool was introduced and evaluated as an offline pipeline for large-scale code corpus cleaning. 
Extending this prior work, the present paper focuses on an IDE-native implementation of \tool that translates SonarQube for IDE diagnostics into interactive, in-place refactoring actions within Visual Studio Code.

\noindent \textbf{Demo Video:} \url{https://www.youtube.com/watch?v=X5BXBqTfmWk}

\noindent \textbf{Plugin download:} \url{https://github.com/Tdcq14/vscode-smellcc.git}
\end{abstract}





\maketitle

\section{INTRODUCTION}
In modern software development, code smells—often manifested as suboptimal design choices like long methods or dead code—represent a silent but significant threat to software maintainability. 
While they do not alter program correctness, these violations of design fundamentals accumulate as technical debt, progressively degrading code quality~\cite{huang2018identifying, gao2021automating, wang2024just, gao2024automating, yan2026evolving}. 
Developers in practice are often constrained by tight release schedules and/or limited resources, forcing them to prioritize feature delivery over code hygiene. 
This leaves developers with little bandwidth to manually identify and sanitize these design flaws. 
Consequently, code smells often remain in the codebase, leading to a steady erosion of software quality and an increase in maintenance costs over time.

\begin{figure*}
    \centering
    \includegraphics[width=\textwidth]{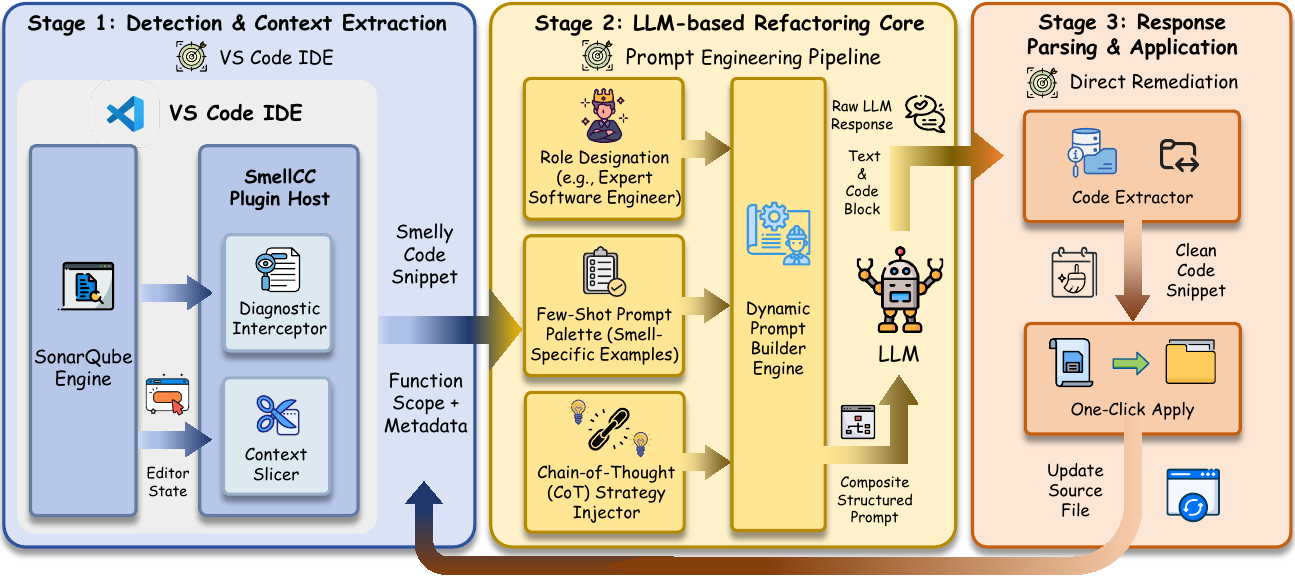}
    \caption{The Overall Framework of Our Approach}
    \label{fig:overview}
\end{figure*}  

In light of these practical limitations, researchers have proposed automated tools for software maintenance. 
% automated support becomes indispensable for maintaining code quality. 
% However, within the current IDE ecosystem, the quality assurance landscape remains fragmented, typically addressing only isolated aspects of the problem. 
RefactorInsight~\cite{kurbatova2021refactorinsight}, for instance, analyzes change history to identify ``hotspots'', effectively prioritizing when to refactor but offering no mechanism to execute the changes. 
Similarly, static analysis plugins of SonarQube~\cite{sonarlint}, while they integrate seamlessly into the IDE to identify issues with high precision, they function largely as passive alert systems.  
They excel at pointing out ``where'' the problem is but lack the capability to resolve ``how'' to fix it, especially for complex logic, thereby leaving the burden of refactoring entirely on the developer. 
To address this gap, early automated tools such as Rope~\cite{rope} for Python, CppRefactory~\cite{cpprefactory} for C++, and JDeodorant~\cite{tsantalis2008jdeodorant} for Java, pioneered rule-based refactoring. 
However, their reliance on predefined rules often limits them to specific crafted patterns, lacking the semantic understanding to address semantic issues. 
Recently, research has pivoted toward deep learning to overcome these limitations: 
Ma et al.~\cite{ma2023pre} employed CodeT5 to target Feature Envy, Tufano et al.~\cite{tufano2019learning} treated refactoring as a neural translation task, and EM-Assist~\cite{wadhwa2024core} leveraged LLMs specifically for Long Parameter Lists. 
While these studies validated the feasibility of LLM-based refactoring, they remain restricted in scope, targeting singular smell types or functioning as empirical explorations. 
Consequently, there is currently no unified tool that organically combines the precision of industry-standard detection with a capability to automatically refactor diverse code smells.

To bridge this gap, we present \tool, a Visual Studio Code plugin designed to automatically refactor Python code smells. 
Fundamentally, our \tool is engineered to augment and complete the capability of SonarQube by triggering automated refactoring after a smell is detected.
Our approach leverages an LLM-based pipeline employing Chain-of-Thought and few-shot learning techniques to ensure that the refactored code is not only clean but also context-aware and behavior-preserving.
\tool specifically targets the ten most frequent code smells identified in our prior study~\cite{10.1145/3793252}—ranging from syntactic issues like \textit{Dead Code} to semantic challenges like \textit{High Cognitive Complexity}—and provides in-place, one-click refactoring. 
This allows developers to instantly sanitize their codebase without context switching, effectively preventing the accumulation of technical debt during development.

%Early seminal works, such as JDeodorant~\cite{tsantalis2008jdeodorant}, pioneered rule-based refactoring for Java, while subsequent tools like RefactorInsight~\cite{kurbatova2021refactorinsight} and Designite ~\cite{sharma2016designite} shifted focus toward mining refactoring history and detecting architectural decay. 
%In the industrial realm, platforms like CodeScene\footnote{\url{https://codescene.com}} introduced behavioral analysis to prioritize technical debt. 
%Furthermore, many tools are difficult to integrate seamlessly into the developer's daily workflow. 
\section{APPROACH}
%Inspired by the great potential of LLMs for code comprehension and code generation, we designed an LLM-based tool, named \tool, for cleaning code smells. 
Fig.~\ref{fig:overview} illustrates the overall framework of \tool, which consists of three interconnected phases: the detection and context extraction phase, the LLM-based refactoring core phase, and the response parsing and application phase. 
It implements a closed-loop refactoring pipeline integrated directly into VS Code.

\subsection{Smell Detection and Context Extraction}
To ensure high-precision identification without redundant computation, \tool functions as an intelligent downstream consumer of the SonarQube engine.
The process begins with the Diagnostic Interceptor, which continuously monitors the IDE's diagnostic collection and selectively captures telemetry data—including rule keys and cursor coordinates—for the targeted top-10 smell types. 
Subsequently, the Context Slicer performs a structural analysis of the Abstract Syntax Tree (AST) to expand the selection from the error location to its enclosing semantic scope (e.g., the full function body), which ensures that we extract the complete code context necessary for the LLM to understand the logic and perform a valid refactoring.

\subsection{LLM-based Code Smell Refactoring}
The underlying approach of \tool is prompt engineering, i.e., using natural language to guide LLMs to complete specific tasks. 
Since LLMs are not designed for refactoring purposes, we leverage \textbf{prompt role designation}, \textbf{chain-of-thought reasoning}, and \textbf{few-shot learning} techniques, which are widely utilized and verified in previous studies~\cite{xue2024selfpico, mai2024human, mai2025towards, dai2024mpcoder},
to harness the LLMs' knowledge for automated code smell refactoring.

\textbf{Prompt Role Designation.} 
In prompt engineering, role designation is a method where LLMs are designated a role for solving a specific task. 
Assigning a role to the LLM provides it with the problem context and leads to more accurate and relevant responses. 
In this study, we design the role of the LLM as ``\textit{an expert software engineer}''. 
After assigning the role, we inform the LLM with our task as follows: ``\textit{Your task is to refactor code to eliminate code smell while keeping the code functionality.}'' 
This can elicit the programming knowledge of the LLM for performing refactoring task. 

\textbf{Chain-of-Thought Reasoning.} 
Chain-of-Thought reasoning is an important strategy for prompt engineering. 
It enables LLMs to split a complicated task into several relatively simple steps and generate a series of intermediate outputs that lead to a reasonable result. 
Our task of cleaning different code smells requires logical thinking to understand the code and a coherent series of intermediate steps to clean smells. 
In order to construct a reasonable CoT, we invite two developers, with over 6 and 9 years of Python programming experience, to manually fix the 10 different code smells respectively and write down their core steps as their chain-of-thought reasoning steps. 
Afterwards, the first author discussed with the two developers to summarize the final steps for each code smell. 
We take the \textit{Collapsible If Statements} smell as an example to show the chain-of-thought reasoning process: (1) Step1: Understanding the task requirement; (2) Step2: Analyze the content of two conditions; 
(3) Step3: Determine the conjunction; (4) Combine the conditions.  
Overall, this step-by-step thinking guides the LLM to clean smells in a manner similar to a developer. 
The detailed CoT for other code smells can be found in our replication package~\cite{replication_package}.

\textbf{Few-shot Learning.} 
With the increasing ability of LLMs, in-context learning has been widely adopted as zero-shot learning and few-shot learning. 
Few-shot learning is utilized to augment the context with a few examples of desired inputs and outputs, which helps the model elicit specific knowledge and abstractions needed to complete the task. 
Regarding the \textit{Collapsible If Statements} smell, we give an example of the code smell input (e.g., \texttt{if condition1: if condition2: \#code}) and the expected code smell output (e.g., \texttt{if condition1 and condition2: \#code}). 
With the help of these representative examples, the LLM can enhance its understanding of the target code smell and its effectiveness for refactoring. 

\textbf{Prompt Assembly and Inference.}
Finally, the Dynamic Prompt Builder Engine synthesizes these strategic components with the detected code context to construct a Composite Structured Prompt. 
This unified instruction drives the LLM to generate a response comprising both the step-by-step rationale and the refactored code block, ready for extraction in the next stage.

\subsection{Response Parsing and Application}
In the final stage, the Code Extractor isolates the refactored logic from the LLM's conversational output to produce a Clean Code Snippet. This snippet is queued for the One-Click Apply mechanism, which allows the developer to instantly inject the optimized code into the editor, physically replacing the detected smell and resolving the technical debt in situ.
\section{IMPLEMENTATION DETAILS}
We implement \tool in the form of a Visual Studio Code plugin. 
%The source code can be found in our Github repository\footnote{\url{https://github.com/Tdcq14/vscode-smellcc.git}}.

\begin{figure}
    \centering
    \includegraphics[width=0.48\textwidth]{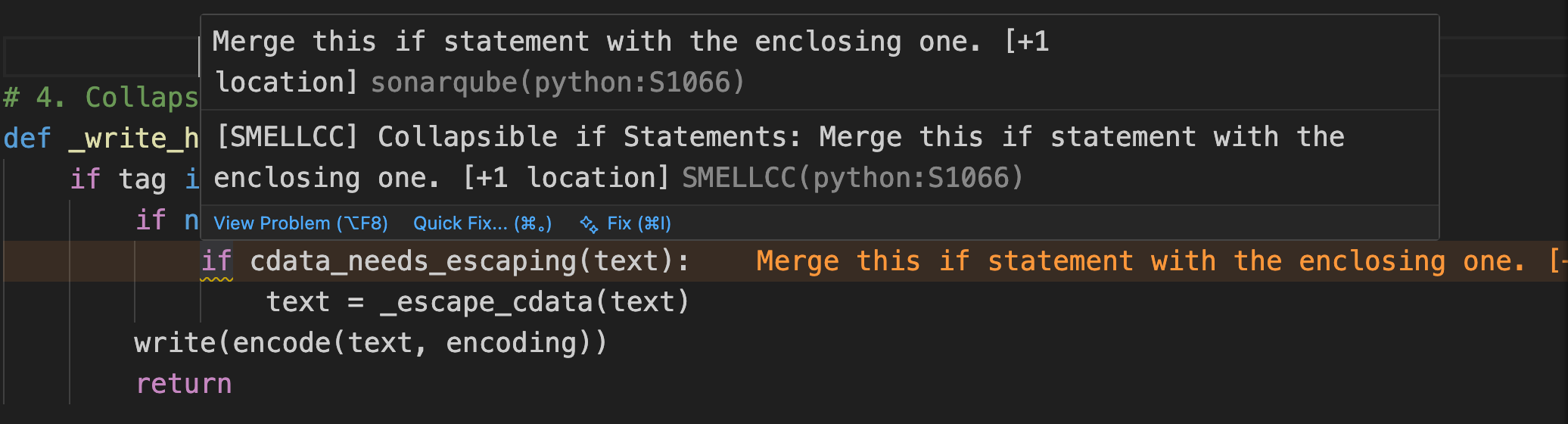}
    \caption{Real-time detection and tooltip explanation of a code smell in VS Code}
    \label{fig:detected_smell}
\end{figure} 

\begin{figure}
    \centering
    \includegraphics[width=0.48\textwidth]{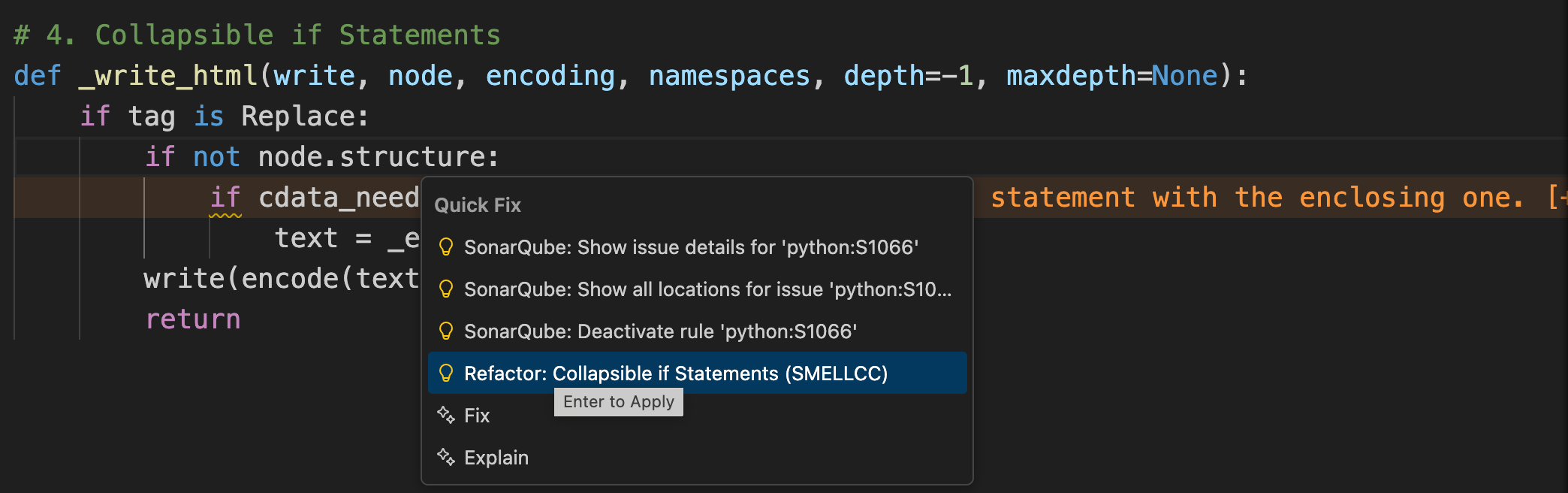}
    \caption{The interactive Quick Fix menu allowing developers to trigger \tool for in-situ refactoring}
    \label{fig:how_to_refactor}
\end{figure} 

\begin{table*} 
\small
\setlength{\abovecaptionskip}{0pt}
  \centering
  \caption{Code Smell Refactoring and Code Correctness Testing of 50 projects}
  \label{tab:verify_functional_correctness}
  \begin{tabularx}{0.66\textwidth}{cccccccc}
    \toprule
\multirow{2}{*}{Code Smell Type} & \multicolumn{3}{c}{Code Smell Refactoring} & \multicolumn{3}{c}{Code Correctness Testing} \\
    \cmidrule(lr){2-4} \cmidrule(lr){5-7}
     & \#Before & \#After & \#Cleaning(\%) & \#Before & \#After & \#Accuracy(\%) \\
    \midrule
    \textit{Naming Convention}  & 11,579 & 62 & 99.5 & 11,579 & 10,578 & 91.4 \\
    \textit{High Cognitive Complexity} & 1,840 & 174 & 90.5 & 1,840 & 1,627 & 88.4 \\
    \textit{Commented Code} & 1,100 & 0 & 100.0 & 1,100 & 1,100 & 100.0 \\
    \textit{Long Parameter List} & 655 & 256 & 60.9 & 655 & 474 & 72.4 \\
    \textit{Collapsible if Statements} & 648 & 0 & 100.0 & 648 & 648 & 100.0 \\
    \textit{Empty Nested Code Blocks} & 109 & 16 & 85.3 & 109 & 107 & 98.2 \\
    \textit{Dead Code} & 37 & 0 & 100.0 & 37 & 37 & 100.0 \\
   \textit{Self-assigned Variables} & 13 & 0 & 100.0 & 13 & 13 & 100.0 \\
    \textit{Identical Expressions} & 8 & 0 & 100.0 & 8 & 8 & 100.0 \\
    \textit{Return and Yield} & 5 & 0 & 100.0 & 5 & 5 & 100.0 \\
    \midrule
    All & 15,994 & 508 & 96.8 & 15,994 & 14,597 & 91.3 \\
    \bottomrule                                
  \end{tabularx}                              
\end{table*}

\textbf{Tool Architecture.} 
\tool comprises a diagnostic interceptor, a context slicer, a dynamic prompt builder engine, an LLM client, a response parser, and an edit provider. 
For each code smell, the diagnostic interceptor first consumes the rule identifier and source range published by SonarQube for IDE and determines whether the issue belongs to one of the ten supported smell categories. 
Then the context slicer expands the reported location to its enclosing function scope. 
After that, the dynamic prompt builder engine selects the corresponding smell-specific template inherited from our prior work~\cite{10.1145/3793252} and instantiates it with the extracted code and diagnostic metadata. 
After asynchronous LLM inference, the response parser further extracts the generated code, and the edit provider applies it to the original source range through the VS Code editing API.

\textbf{Configuration and Deployment.} 
\tool is distributed as a VS Code extension package through GitHub Releases and requires SonarQube for IDE, JRE 17+, and Internet access to an LLM service. 
Through the VS Code settings interface, developers provide an API key and may configure the API endpoint and model, with DeepSeek-Coder used by default.
For LLM, we chose the DeepSeek-Coder model, which has proven to have excellent abilities in coding tasks~\cite{zhu2024deepseek}, achieving a close and comparable performance with GPT-4o. 
Moreover, the DeepSeek-Coder model is rather cost-effective, compared with GPT-4o, the input token costs just 2.8\% and the output token costs just 1.9\% of the GPT-4o. 
Considering the high volume of requests anticipated in a daily coding workflow, we chose DeepSeek-Coder for its powerful code generation capabilities and affordability, making it a sustainable choice for continuous automated refactoring. 
When using LLMs, the parameter of temperature is a pivotal setting that governs the randomness of the model's output. 
In this study, we set the temperature at 0 by a series of pilot experiments. 
According to the DeepSeek documentation, the \textit{max\_tokens} can reach 8,192. 
In this study, when handling code smells such as \textit{Long Parameter List}, we set the \textit{max\_tokens} to 8,192 to deal with the large number of tokens. 
For other cases, it was set to 2,048 to save time and resources.

\textbf{User Interface.}
Fig.~\ref{fig:detected_smell} and Fig.~\ref{fig:how_to_refactor} showcase the user interface of \tool. 
To initiate the workflow, developers need to install the \tool extension alongside the prerequisite SonarQube for IDE plugin first. 
The subsequent usage cycle consists of two primary steps: detection and interactive remediation. 
Once these extensions are installed, the detection mechanism operates automatically in the background. 
When a code smell is identified—such as the \textit{Collapsible If Statements} shown in Fig.~\ref{fig:detected_smell}—the editor highlights the problematic code with a diagnostic underline. 
Hovering over this indicator reveals a detailed tooltip that specifies the smell type and explains the underlying issue. 
To eliminate the identified smell, the developer simply engages the standard ``Quick Fix'' mechanism. 
As illustrated in Fig.~\ref{fig:how_to_refactor}, this action summons a menu featuring the specific remediation command: ``Refactor: [Smell Type] (SMELLCC)''. 
Selecting this option triggers the refactoring pipeline, which automatically replaces the smelly code snippet with a clean version in situ.
In this process, \tool invokes the LLM asynchronously, allowing the editor to remain responsive while the request is processed. 
Once a valid response is received, \tool parses the generated code and applies it to the target code for immediate inspection. The edit is recorded as a single editor operation and can be reverted using VS Code’s standard Undo command. 
If the request or response parsing fails, the extension would report the error and leave the source code unchanged.

\section{EVALUATION}
We evaluate the performance of \tool through a quantitative analysis centered on two aspects: the success rate of code smell refactoring (Effectiveness) and the pass rate of regression tests on the refactored code (Correctness).

\noindent\textbf{Data Collection.} 
We manually curated a dataset of 50 projects from \texttt{CodeSearchNet}-Python. 
To ensure diverse code smell coverage, we prioritized projects with high function counts. 
Each candidate underwent strict manual verification of its testing infrastructure (e.g., presence of test*.py or PyTest); only projects with successfully executable test suites were retained. 
The final dataset consists of 50 fully configured, verifiable projects ready for evaluation.

\noindent\textbf{Experiment Setup.} 
First, to assess refactoring effectiveness, we employed SonarQube to quantify the reduction in the top-10 code smell types across the dataset before and after remediation. 
%(which shares the same underlying analysis engine as SonarLint)
Then, to verify functional correctness, we executed regression tests on the refactored code, calculating the success rate to ensure that the original program behavior was strictly preserved.

\noindent\textbf{Experiment Result.}
Table~\ref{tab:verify_functional_correctness} presents the results of code smell elimination and functional correctness testing of 50 projects.
The experimental results show that SonarQube detected 15,994 code smells from the above projects, and \tool \textbf{successfully removed 96.8\%} of them (15,486 out of 15,994). 
By running unit tests, we found that \textbf{91.3\%} (14,597 out of 15,994) of the refactored code maintained the same functionality as the original one. 
This high success rate validates the efficacy of our \tool in handling diverse technical debts.
From the perspective of different code smell types, several points stand out:

\faThumbsUp~\textbf{Perfect Refactoring Cases}: 
From the table, we can see that \tool shows perfect (i.e., 100\%) effectiveness and correctness for six smells: \textit{Commented Code}, \textit{Collapsible if Statements}, \textit{Dead Code}, \textit{Self-assigned Variables}, \textit{Identical Expressions}, and \textit{Return and Yield}. 
These code smells are relatively straightforward to detect and refactor by syntactic patterns or pre-defined rules; the perfect performance in these cases highlights \tool’s strong capability in handling structurally simple yet common syntactic code issues.

\faThumbsUp~\textbf{High-Performance Cases}: 
\tool is highly effective for cleaning the following three code smells: \textit{Naming Convention} (99.5\% effectiveness and 91.4\% correctness), \textit{High Cognitive Complexity} (90.5\% effectiveness and 88.4\% correctness), and \textit{Empty Nested Code Blocks} (85.3\% effectiveness and 98.2\% correctness). 
These smell types require a deeper understanding of code semantics and developer’s intent, which goes beyond the surface-level pattern matching. 
For instance, addressing \textit{High Cognitive Complexity} requires the model to recognize logical structures, abstract control flow, and restructure the code while preserving its behavior. 
The success of \tool on these cases demonstrates its effectiveness in capturing code context and adhering to structural constraints. 
It is able to identify complex syntactic patterns and generate standard-compliant edits. 

\faThumbsDown~\textbf{Challenging Cases}: 
The \textit{Long Parameter List} code smell demonstrates the lowest refactoring effectiveness among all ten categories, achieving only a 60.9\% cleaning rate with 72.4\% correctness. 
This relatively poor performance reveals the limitation of \tool in handling smells that require more context information and global reasoning, suggesting future improvements through providing more dependency information, extending the context window and employing multi-agent collaboration across multiple functions and files.

\textbf{Implications for Reliable Refactoring.}
The experimental results show that \tool is more reliable for localized, syntax-oriented smells than for refactorings involving broader dependencies, such as \textit{High Cognitive Complexity} and \textit{Long Parameter List}. 
This suggests that improving reliability requires both stronger behavioral constraints and richer program context. 
Future versions could incorporate dependency-aware context, reasoning steps that examine call sites and side effects, and few-shot examples matched to the target code structure. 
Combined with post-generation validation, these improvements could reduce behavioral regressions in complex refactorings.

\section{CONCLUSION AND FUTURE WORK}
In this paper, we present \tool, an IDE-native automated refactoring tool that bridges the gap between code smell detection and refactoring.
By synergizing the precision of static analysis (via SonarQube) with the generative capabilities of LLMs (DeepSeek-Coder), \tool offers a streamlined, one-click solution for managing code smells. 
Our evaluation demonstrates its efficacy, achieving a 96.8\% remediation rate and 91.3\% functional correctness across 50 Python projects. 
In the future, we plan to extend \tool by supporting more programming languages (e.g., Java), more IDEs (e.g., IntelliJ IDEA), and configurable LLM backends, while incorporating project-level context to address complex architectural smells in large-scale development environments.

\section*{Acknowledgment}
This research is supported by the National Science Foundation of China (No. 62572322).  
This research is partially sponsored by the CCF-Huawei Populus Grove Fund and Tencent Rhino-Bird Fund. 
We also thank the anonymous reviewers for their insightful comments and suggestions.

\balance
\bibliographystyle{ACM-Reference-Format}
\bibliography{software}

@inproceedings{tsantalis2008jdeodorant,
  title={JDeodorant: Identification and removal of type-checking bad smells},
  author={Tsantalis, Nikolaos and Chaikalis, Theodoros and Chatzigeorgiou, Alexander},
  booktitle={2008 12th European conference on software maintenance and reengineering},
  pages={329--331},
  year={2008},
  organization={IEEE}
}

@inproceedings{kurbatova2021refactorinsight,
  title={Refactorinsight: Enhancing ide representation of changes in git with refactorings information},
  author={Kurbatova, Zarina and Kovalenko, Vladimir and Savu, Ioana and Brockbernd, Bob and Andreescu, Dan and Anton, Matei and Venediktov, Roman and Tikhomirova, Elena and Bryksin, Timofey},
  booktitle={2021 36th IEEE/ACM International Conference on Automated Software Engineering (ASE)},
  pages={1276--1280},
  year={2021},
  organization={IEEE}
}

@misc{sonarlint,
  author = {{SonarSource}},
  title = {{SonarLint: IDE static code analysis}},
  year = {2025},
  url = {https://marketplace.visualstudio.com/items?itemName=SonarSource.sonarlint-vscode}
}

@misc{cpprefactory,
  title = {CPPRefactory},
  author = {blep, jejudd, svenreichard},
  year = {2013},
  howpublished = {\url{https://sourceforge.net/projects/cpptool/}},
  note = {Last accessed: September 9, 2024}
}

@misc{rope,
  title = {Rope: A Python Refactoring Library},
  author = {Ali Gholami Rudi and Lie Ryan},
  year = {2024},
  howpublished = {\url{https://github.com/python-rope/rope}},
  note = {Last accessed: September 9, 2024}
}

@inproceedings{ma2023pre,
  title={Pre-trained Model Based Feature Envy Detection},
  author={Ma, Wenhao and Yu, Yaoxiang and Ruan, Xiaoming and Cai, Bo},
  booktitle={2023 IEEE/ACM 20th International Conference on Mining Software Repositories (MSR)},
  pages={430--440},
  year={2023},
  organization={IEEE}
}

@inproceedings{tufano2019learning,
  title={On learning meaningful code changes via neural machine translation},
  author={Tufano, Michele and Pantiuchina, Jevgenija and Watson, Cody and Bavota, Gabriele and Poshyvanyk, Denys},
  booktitle={2019 IEEE/ACM 41st International Conference on Software Engineering (ICSE)},
  pages={25--36},
  year={2019},
  organization={IEEE}
}

@article{wadhwa2024core,
  title={CORE: Resolving Code Quality Issues using LLMs},
  author={Wadhwa, Nalin and Pradhan, Jui and Sonwane, Atharv and Sahu, Surya Prakash and Natarajan, Nagarajan and Kanade, Aditya and Parthasarathy, Suresh and Rajamani, Sriram},
  journal={Proceedings of the ACM on Software Engineering},
  volume={1},
  number={FSE},
  pages={789--811},
  year={2024},
  publisher={ACM New York, NY, USA}
}

@article{zhu2024deepseek,
  title={DeepSeek-Coder-V2: Breaking the Barrier of Closed-Source Models in Code Intelligence},
  author={Zhu, Qihao and Guo, Daya and Shao, Zhihong and Yang, Dejian and Wang, Peiyi and Xu, Runxin and Wu, Y and Li, Yukun and Gao, Huazuo and Ma, Shirong and others},
  journal={arXiv preprint arXiv:2406.11931},
  year={2024}
}

@misc{replication_package,
  title = {Our replication package},
  year = {2024},
  howpublished = {https://zenodo.org/records/17578401}
}

@article{10.1145/3793252,
author = {Xue, Zhipeng and Zhang, Xiaoting and Gao, Zhipeng and Hu, Xing and Gao, Shan and Xia, Xin and Li, Shanping},
title = {Clean Code, Better Models: Enhancing LLM Performance with Smell-Cleaned Dataset},
year = {2026},
publisher = {Association for Computing Machinery},
address = {New York, NY, USA},
issn = {1049-331X},
url = {https://doi.org/10.1145/3793252},
doi = {10.1145/3793252},
note = {Just Accepted},
journal = {ACM Trans. Softw. Eng. Methodol.}
}

@inproceedings{gao2021automating,
  title={Automating the removal of obsolete TODO comments},
  author={Gao, Zhipeng and Xia, Xin and Lo, David and Grundy, John and Zimmermann, Thomas},
  booktitle={Proceedings of the 29th ACM Joint Meeting on European Software Engineering Conference and Symposium on the Foundations of Software Engineering},
  pages={218--229},
  year={2021}
}

@article{wang2024just,
  title={Just-in-time todo-missed commits detection},
  author={Wang, Haoye and Gao, Zhipeng and Hu, Xing and Lo, David and Grundy, John and Wang, Xinyu},
  journal={IEEE Transactions on Software Engineering},
  volume={50},
  number={11},
  pages={2732--2752},
  year={2024},
  publisher={IEEE}
}

@article{gao2024automating,
  title={Automating todo-missed methods detection and patching},
  author={Gao, Zhipeng and Su, Yanqi and Hu, Xing and Xia, Xin},
  journal={ACM Transactions on Software Engineering and Methodology},
  volume={34},
  number={1},
  pages={1--28},
  year={2024},
  publisher={ACM New York, NY, USA}
}

@article{yan2026evolving,
  title={Evolving Trends in Cleanliness of Open Source Projects},
  author={Yan, Dapeng and Yang, Wenjie and Gao, Zhipeng and Liu, Kui and Cai, Zhikuang and Xie, Xiaoyuan and Liu, Zhiming},
  journal={ACM Transactions on Software Engineering and Methodology},
  year={2026},
  publisher={ACM New York, NY}
}

@article{huang2018identifying,
  title={Identifying self-admitted technical debt in open source projects using text mining},
  author={Huang, Qiao and Shihab, Emad and Xia, Xin and Lo, David and Li, Shanping},
  journal={Empirical Software Engineering},
  volume={23},
  number={1},
  pages={418--451},
  year={2018},
  publisher={Springer}
}

@inproceedings{xue2024selfpico,
  title={Selfpico: Self-guided partial code execution with llms},
  author={Xue, Zhipeng and Gao, Zhipeng and Wang, Shaohua and Hu, Xing and Xia, Xin and Li, Shanping},
  booktitle={Proceedings of the 33rd ACM SIGSOFT International Symposium on Software Testing and Analysis},
  pages={1389--1401},
  year={2024}
}

@article{mai2024human,
  title={Are human rules necessary? generating reusable apis with cot reasoning and in-context learning},
  author={Mai, Yubo and Gao, Zhipeng and Hu, Xing and Bao, Lingfeng and Liu, Yu and Sun, JianLing},
  journal={Proceedings of the ACM on Software Engineering},
  volume={1},
  number={FSE},
  pages={2355--2377},
  year={2024},
  publisher={ACM New York, NY, USA}
}

@inproceedings{mai2025towards,
  title={Towards better answers: Automated stack overflow post updating},
  author={Mai, Yubo and Gao, Zhipeng and Wang, Haoye and Bi, Tingting and Hu, Xing and Xia, Xin and Sun, Jianling},
  booktitle={2025 IEEE/ACM 47th International Conference on Software Engineering (ICSE)},
  pages={591--603},
  year={2025},
  organization={IEEE}
}

@inproceedings{dai2024mpcoder,
  title={Mpcoder: Multi-user personalized code generator with explicit and implicit style representation learning},
  author={Dai, Zhenlong and Yao, Chang and Han, WenKang and Yuanying, Yuanying and Gao, Zhipeng and Chen, Jingyuan},
  booktitle={Proceedings of the 62nd Annual Meeting of the Association for Computational Linguistics (Volume 1: Long Papers)},
  pages={3765--3780},
  year={2024}
}

\end{document}